\documentclass[11pt,preprint2]{emulateapj}

\usepackage {pstricks}

\shorttitle{Presupernova Instabilities}
\shortauthors{Smith \& Arnett}

\newcommand{\etal} { et~al.\ }  
\newcommand{\sol}{$M_\odot$}
\def \nuc#1#2{\relax\ifmmode{}^{#1}{\protect\text{#2}}\else${}^{#1}$#2\fi}
\def \msol#1{\relax$#1\,M_\odot\/$}

\begin{document}

\title{Preparing for an Explosion: Hydrodynamic Instabilities and
  Turbulence in Presupernovae}

\author{Nathan Smith\altaffilmark{1,2}}

\author{W. David Arnett\altaffilmark{1,3,4,5}}

\altaffiltext{1}{Steward Observatory, University of Arizona, 
933 N. Cherry Avenue, Tucson AZ 85721}

\altaffiltext{2}{Email: nathans@as.arizona.edu}

\altaffiltext{3}{Email: darnett@as.arizona.edu}

\altaffiltext{4}{Kavli Institute for Theoretical Physics, University of California,
Santa Barbara CA}

\altaffiltext{5}{Aspen Center for Physics, Aspen CO}

\begin{abstract}

  Both observations and numerical simulations are discordant with
  predictions of conventional stellar evolution codes for the latest
  stages of a massive star's life prior to core collapse.  The most
  dramatic example of this disconnect is in the eruptive mass loss
  occurring in the decade preceding Type IIn supernovae (SNe).  We
  outline the key empirical evidence that indicates severe pre-SN
  instability in massive stars, and we suggest that the chief reason
  these outbursts are absent in stellar evolution models may lie in
  the treatment of turbulent convection in these codes.  The mixing
  length theory that is used ignores finite amplitude fluctuations in
  velocity and temperature, and their nonlinear interaction with
  nuclear burning.  Including these fluctuations is likely to give
  rise to hydrodynamic instabilities in the latest burning sequences,
  which prompts us to discuss a number of far-reaching implications
  for the fates of massive stars.  In particular, we explore
  connections to enhanced pre-supernova mass loss, unsteady nuclear
  burning and consequent eruptions, swelling of the stellar radius
  that may trigger violent interactions with a companion star, and
  potential modifications to the core structure that could
  dramatically alter calculations of the core-collapse explosion mechanism
  itself.  These modifications may also impact detailed
  nucleosynthesis and measured isotopic anomalies in meteorites, as
  well as the interpretation of young core-collapse SN remnants.
  Understanding these critical instabilities in the final stages of
  evolution may make possible the development of an early warning
  system for impending core collapse, if we can identify their
  asteroseismological or eruptive signatures.

\end{abstract}

\keywords{convection --- instabilities --- stars: supernovae --- stars:
  mass loss --- meteorites --- nucleosynthesis}

\section{Introduction}

A significant subset of supernovae (SNe) appear to have suffered heavy
and episodic pre-SN mass loss (Smith et al.\ 2011a; Smith 2014).  This
has not been explained by standard stellar evolutionary models, nor
have the effects of episodic mass loss been included in them (see
Langer [2012] and Maeder \& Meynet [2000]).  In part this may be
because stellar evolution calculations are one dimensional (1D), have
large time steps which damp dynamic effects, and many do not even
progress past carbon burning. Such an approach may be valid for lower
masses, but at higher masses several physical processes conspire to
cause problems: (1)
luminosities near the Eddington limit, (2) adiabatic exponent $\gamma$
approaching $4/3$ (or less) due to formation of electron-positron
pairs, (3) complex and unstable nuclear burning [quasi-equilibrium
(QSE) and nuclear statistical equilibrium (NSE) during oxygen and
silicon burning], and (4) accelerated evolution due to copious energy
loss by neutrinos.  All these effects can become important and
problematic if the stars suffer hydrodynamic instability, and could
lead to significant errors in models of pre-SN stars.

There is empirical evidence of pre-SN hydrodynamic instability in both
evolved massive stars in the Milky Way and Local Group, and in the
dense circumstellar material (CSM) that must exist around Type~IIn SNe
and related events (see \S2 for details).  In the Milky Way and other
local galaxies, the massive stars which are most notoriously observed
to experience violent episodic mass loss are the luminous blue
variables (LBVs; Conti 1984; Humphreys \& Davidson 1994; Smith et al.\
2004, 2011b; Clark et al.\ 2005).  This class of objects unifies a
number of different specific types of blue supergiant variable stars,
not all of which are necessarily in identical evolutionary phases, and
most of which have persisted without a SN explosion for centuries.
Most relevant are the so-called giant eruptions of LBVs, exemplified
by the 19th century Great Eruption of $\eta$ Carinae (see below),
although $\eta$ Car is perhaps the most extreme example of the
phenomenon.

Extragalactic Type~IIn SNe, named for their narrow H emission lines,
are unique among SNe because they require CSM that is massive enough
to decelerate the fast SN ejecta, allowing radiation from slow, heated
material to dominate the observed spectrum.  In order for the shock
interaction to occur simultaneously and veil the broad lines from the
underlying SN photosphere, this CSM must reside very close to the star
(within a few 10$^{15}$ cm) and its ejection by the star must
therefore have been synchronized to occur within only a few years of
core collapse.  This synchronicity, fully appreciated only in the last
decade or so, hints at some as-yet unidentified phenomenon associated
with the final nuclear burning phases in the star's life.  We explore
this topic in more detail in this paper.

It has been proposed that eruptive LBV-like events are connected to
SNe~IIn (Smith et al.\ 2007, 2008, 2010a; Gal-Yam et al.\ 2007), but
this has been controversial.  It is in direct conflict with stellar
evolution models for massive stars, all of which currently assume that
massive stars with $M_{\rm ZAMS} \ge 40 \, M_{\odot}$ \, at near-Solar
metallicity will shed their H envelopes via steady winds to make
Wolf-Rayet (WR) stars before core collapse as SNe of Type
Ibc.\footnote{This conflict with stellar evolution models persists
  even if we admit that the ``LBV-like'' pre-SN eruptions inferred for
  SNe~IIn may not necessarily be from the same unstable stars that
  make up nearby LBVs (many of which have not exploded within
  centuries of a past giant LBV eruption). The conflict is mainly in
  the substantial mass of H left in the envelopes of these very
  massive stars at the time of death.} With currently adopted
mass-loss rates (see Smith 2014) models predict that the most massive
stars shed all their H envelope 0.5-1 Myr before exploding.  However,
a direct LBV/SN connection was recently confirmed in the remarkable
case of SN~2009ip, which was a very massive star (50-80 \sol) that was
being studied in detail as an LBV with multiple eruptions (see Smith
et al.\ 2010b; Foley et al.\ 2011).  While under study, this same
eruptive star was then seen to explode as a Type~IIn SN in 2012
(Mauerhan et al.\ 2013), providing the best observed case yet of
unstable pre-SN evolution.

In this paper we take a closer look at the latest stages of
pre-core-collapse evolution and discuss some intriguing possibilities
for how these pre-SN events may occur.  After summarizing the key
observational evidence for episodic pre-SN mass loss in \S2, we
discuss the detailed treatment of convection in stellar interiors in
\S3, and we then discuss a number of implications for the resulting
mass loss and stellar structure in late evolution in \S4.

\section{Relevant Observational Evidence \label{section2}}

In the past decade, empirical evidence has made it very clear that
massive stars can experience violent episodic mass-loss events in
their late evolutionary phases.  Eruptive mass loss may be more
important than steady winds in the integrated mass shed by the star in
its lifetime (Smith \& Owocki 2006), but the current generation of
stellar evolution models does not account for it.  Our aim here is to
provide only a brief overview of the different lines of observational
evidence; a more thorough discussion of pre-SN mass loss will appear
in a forthcoming review (Smith 2014).  We wish to diagnose the
physical mechanisms that lead to this observed mass loss, and to this
end, we first summarize the most important aspects of the relevant
empirical evidence, including the types of progenitor stars that
undergo these pre-SN outbursts, as well as timescales, mass budget,
and energy budget.

\subsection{LBVs} 

The LBVs are a rather diverse class, consisting of a wide range of
irregular variable phenomena associated with evolved massive stars
(Conti 1984; Humphreys \& Davidson 1994; van Genderen 2001; Smith et
al.\ 2004, 2011b), excluding those seen in cool supergiants.  The most
relevant variability attributed to LBVs is the so-called ``giant
eruption'', in which stars are observed to dramatically increase their
luminosity for months to years, accompanied by severe mass loss.  The
best studied example is the Galactic object $\eta$ Carinae, providing
us with its historically observed light curve (Smith \& Frew 2011;
Frew 2004), as well as its complex ejecta that contain 10-20 \sol\ and
$\sim$10$^{50}$ ergs of kinetic energy (Smith et al.\ 2003).  Aside
from the less well-documented case of P Cygni's 1600 AD eruption, our
only other examples of LBV-like giant eruptions are in nearby
galaxies.  The associated optical transients are often discovered in
dedicated searches for SNe, and hence, they have sometimes been
referred to more generally as ``SN impostors'', since their connection
to true LBVs is not always clear.  A number of these have been
identified, with peak luminosities similar to $\eta$ Car or less (Van
Dyk \& Matheson 2012; Smith et al.\ 2011b).  Typical bulk expansion
speeds in the ejecta are 100--1000 km s$^{-1}$ (Smith et al.\ 2011b).
LBVs are generally thought to be very massive stars, but their mass
range is known to extend down to 25 \sol\ (Smith et al.\ 2004) and
some of the extragalacitc SN impostors appear to have relatively
low-mass progenitors around 8 \sol\ (e.g., Prieto et al.\ 2008;
Thompson et al.\ 2009).

Although the giant eruptions themselves are rarely observed because
they are infrequent and considerably fainter than SNe, a large number
of LBVs and spectroscopically similar stars in the Milky Way and
Magellanic Clouds are surrounded by massive shell nebulae, indicating
previous eruptions with a range of ejecta masses from 1--20 \sol
(Clark et al.\ 2005; Smith \& Owocki 2006; Wachter et al.\ 2010;
Gvaramadze et al.\ 2010).  Thus, LBV mass loss is inferred to be
important in late evolution of massive stars.

Traditionally, LBVs have been cast as super-Eddington winds driven by
an increase in the star's bolometric luminosity (Humphreys \& Davidson
1994; Shaviv 2000; Owocki et al.\ 2004; Smith \& Owocki 2006), but
there is growing evidence that some of them are explosive hydrodynamic
ejections (see Smith 2008, 2013).  Of course, these two cases
(long-lived super-Eddington winds or sudden hydrodynamic explosions)
may both operate.

\subsection{Type IIn supernovae and pre-SN mass loss}

As noted above, the narrow H lines in SNe~IIn require very dense CSM
ejected shortly before core collapse.  As the SN blast wave expands
into the CSM, the SN ejecta are decelerated and the slow CSM is heated
(e.g., Chugai et al.\ 2004; Chugai \& Danziger 1994).  Through
conservation of momentum, one can infer the mass of CSM required to
decelerate the fast SN ejecta (Smith et al.\ 2010a).  Special cases of
this are the super-luminous SNe~IIn (SN~2006gy, SN~2006tf, SN~2003ma,
etc.)  where the bolometric luminosity of the CSM interaction demands
very large masses up to 20\ \sol\ ejected a few years\footnote{In a
  25\ \sol \ star, the neutrino-cooled stage takes about 300 yr, most
  of which is spent in C burning, with 3 years in Ne burning, 1 year O
  burning, and 4 days Si burning (e.g., \citealt{wda96}).  Woosley et
  al.\ (2002) give similar numbers: C burning is 520 yr (1100 yr), Ne
  burning is 0.9 yr (0.6 yr), O burning is 0.4 yr (0.9 yr), and Si
  burning is 0.7 d (2 d) for an initially 25\ \sol\ (75\ \sol)
  star. \cite{limchi} find similar values. The dramatic
  reduction in time scales as burning advances is due to enhanced
  neutrino cooling at higher temperatures. Mixing algorithms affect
  the precise numbers; higher mass stars evolve faster, and have
  little C to burn. All of these estimates are inaccurate to the
    extent that they are sensitive to empirically determined mass-loss
    rates imposed in stellar evolution models, which do not adequately
    include binary effects (see \citealt{ns14} for a detailed
    discussion). } before core collapse (Smith \& McCray 2007; Smith
et al.\ 2007, 2008, 2010a; Woosley et al.\ 2007; Rest et al.\
2011). For the characteristic Type~IIn spectrum and high luminosity
from CSM interaction to occur immediately after explosion, the CSM
must be very close (within few to several $10^{15}$ cm) of the star;
from the widths of narrow lines in spectra (typically 100-600 km
s$^{-1}$; Smith et al.\ 2008, 2010a; Kiewe et al.\ 2012) the mass
ejection must have occurred just a few years before hand. This is a
strong hint that something violent (i.e., hydrodynamic) may be
happening to these stars during O and Ne burning, so that the star's
outer layers are already affected by the impending core collapse.

\subsection{Earlier mass loss\label{earlier_mdot}} 
Some SNe~IIn (and even some other SN types) show indications of more
distant CSM at radii of 10$^3$ AU ($1.5\times 10^{16}$ cm) or more,
suggesting heavy mass loss that occurred over a longer timescale than
just a few years or a decade (at $100$ km/s it takes 150 years to
reach a thousand AU).  The evidence for this is that some SNe~IIn
remain luminous and show signatures of strong CSM/shock interaction
for years as the shock continues to overtake more distant CSM ejected
in centuries to millennia before the SN.  Many SNe~IIn show bright
infrared (IR) echoes from distant dusty CSM heated by the SN
luminosity, with massive shells again residing at radii of up to a few
light years (Fox et al.\ 2011).  Some non-IIn explosions show IR
echoes and may have dense CSM shells as well; an illustrative example
is SN 1987A, where the SN blast wave began crashing into a CSM ring
located 0.2 pc from the star after a delay of 10 yr. Thus, both
persistent CSM interaction and IR echoes suggest that heavy mass loss
is not limited to only the few years preceding core collapse.  On
timescales of 100-1000 yr or more, this may indicate that enhanced
episodic mass loss may be linked to C burning, and possibly even He
burning.  Binary interaction may play a role.  These more distant
(older) CSM shells are harder to detect, so we do not have a reliable
census of what fraction of all SN progenitors experience this
phenomenon.

\subsection{Progenitor eruptions} 

There have now been two\footnote{ Actually, as our manuscript was
  in the review process, Fraser et al. (2013) reported the direct
  detection of another pre-SN outburst in archival data one year
  before the Type IIn-P event SN~2011ht.}  clear direct detections of
a precursor eruption, seen as a transient source just a few years
before the SN.
SN~2006jc was the first object clearly seen to have a brief outburst 2
years before a SN. The precursor event in 2004 had a peak luminosity
similar to that of $\eta$ Car (absolute magnitude of
$-$14)\footnote{With no bolometric correction this is a luminosity of
  $\sim 10^{41}$ erg s$^{-1}$.}, and was fairly brief (only a few
weeks; Pastorello et al.\ 2007). No spectra were obtained for the
precursor transient source, but the SN explosion 2 years later had
strong narrow emission lines of He, indicating dense CSM
(Pastorello et al.\ 2007; Foley et al.\ 2007).\footnote{The Type Ibn
  spectrum of SN~2006jc is quite significant.  Strong narrow He lines
  (and very weak H lines) indicate a He-rich/H-poor CSM.  This, in
  turn, implies that the progenitor probably had a relatively compact
  stellar radius.  The fact that SN~2006jc exhibited the same type of
  pre-SN eruptions as seen in SNe~IIn indicates that an extended
  stellar radius and encounters with a companion may not be the only
  explanation for precursor outbursts, and that something deeper is at
  work. Woosley et al. (2007) suggested that the precursor
    outburst of SN~2006jc might have been a pulsational pair
    instability eruption.}  

  A much more vivid case was SN~2009ip, mentioned earlier.  It was
  initially discovered and studied in detail as an LBV-like outburst
  in 2009, again with a peak absolute magnitude near $-$14 and a
  spectrum similar to LBVs.  A quiescent progenitor star was detected
  in archival {\it Hubble Space Telescope} ({\it HST}) data taken 10
  yr earlier, which indicated a very massive star of 50-80 \sol \
  (Smith et al.\ 2010b; Foley et al.\ 2011).  The object then
  experienced several similar eruptions over 3 yrs that looked like
  additional LBV eruptions (unlike SN~2006jc, detailed spectra of
  these progenitor outbursts were obtained), culminating in a final SN
  explosion in 2012 (Mauerhan et al.\ 2013; Smith et al. 2014).  The
  SN light curve was double-peaked, with initially faint bump ($-$15
  mag) that had very broad emission lines (probably the SN ejecta
  photosphere), and it rose quickly 40 days later to a peak of $-$18
  mag, when it looked like a normal SN IIn (caused by CSM interaction,
  as the SN crashed into the slow material ejected 1-3 years earlier;
  see Mauerhan et al.\ 2013; Smith et al. 2014).  A number of detailed
  studies of the SN have now been published (Mauerhan et al.\ 2013;
  Prieto et al.\ 2013; Pastorello et al.\ 2013; Fraser et al. 2013;
  Smith et al.\ 2013, 2014; Margutti et al.\ 2014; Ofek et al.\
  2013a).  A related case is the Type IIn SN~2010mc (Ofek et al.\
  2013b), which had a double-peaked light curve that was nearly
  identical to SN~2009ip (Smith et al.\ 2013, 2014), but did not have
  the same extensive pre-SN observations or a detection of precursor
  outbursts.  While there was initially some debate about the nature
  of these objects, their late-time data revealed them to be true
  core-collapse events, and Smith et al.\ (2014) demonstrated that
  their double-peaked light curves were well explained as
  core-collapse SNe from blue supergiants, but with strong CSM
  interaction. 

  Another possible case of a detected pre-SN outburst is the
  historical object SN~1961V (Smith et al.\ 2011b; Kochanek et
  al. 2011); its pre-1961 photometry may have indicated a precursor
  outburst, but it was far less clearly delineated than in SN~2006jc
  and SN~2009ip. In any case, if we take the full class of SNe IIn,
  which represent 8-9\% of all core-collapse SNe in large galaxies
  (Smith et al.\ 2011a), the phenomenon is far too common to be caused
  by the pulsational pair instability that operates in very massive
  stars (as discussed below).

\subsection{Progenitor star detections} 

We now have four directly detected quiescent progenitors of SNe~IIn
with rough mass estimates.  (1) SN~2005gj with an implied initial mass
of roughly 60 \sol \ (Gal-Yam et al.\ 2007; Gal-Yam \& Leonard 2009),
(2) SN~1961V with a very luminous progenitor indicating an initial
mass of order 100-150 \sol \ or more (Smith et al.\ 2011b; Kochanek et al.\
2011), (3) SN~2009ip (50-80 \sol; see above), and (4) SN~2010jl with a
likely progenitor mass of $>30$ \sol \ (Smith et al.\ 2011c).  In this
last case, however, the SN has not yet faded, so the candidate
progenitor might be a massive young cluster; if so it still suggests a
massive star above 30 \sol \ (Smith et al.\ 2011c).  All four cases
suggest progenitor stars that are much more massive than the typical
red supergiant progenitors of SNe II-P (Smartt 2009).  (This does not,
however, preclude the possibility that some lower-mass stars produce
SNe~IIn as well, since the most luminous LBV-like progenitors are the
easiest to detect.)

\subsection{Statistics} \label{epair}

From a volume-limited sample in a survey with controlled systematics,
Type~IIn SNe appear to make up roughly 8-9\% of all core-collapse
events (Smith et al.\ 2011a).  To get a sense of what this might mean
we note that it corresponds roughly to the ratio of the number of
stars in the range 30-100\ \sol \ to the total number in 8-100\ \sol \
(\citealt{wda96}, Table~14.4).  SNe~IIn are the most diverse and poorly
understood of all SN types --- any SN type can appear as a Type IIn if
it has dense CSM. Some SNe~IIn appear to be Type Ia with CSM, and some
appear to be low-energy electron capture SNe from 8-10 \sol\ stars,
but the majority appear to be from more massive LBV-like stars (see,
e.g., \citealt{ns14} and references therein).

These statistics indicate that the precursor eruptions leading to
SNe~IIn are far too common for all of them to be attributable to the
pulsational pair instability (\citealt{wda96}, \S11.7; Woosley et al.\
2007; Heger \& Woosley 2002; Chatzopoulos \& Wheeler 2012), which is
expected to occur only in very rare, very massive stars with initial
masses around 100 \sol \ or more. Some other instability must operate
in most SNe~IIn.

The relative rarity of SNe~IIn is also important: As we consider
implications for the dominant physical mechanism that may lead to
pre-SN eruptions, we must be mindful that the vast majority of SNe
($\sim$90\%) {\it do not} suffer violent precursor outbursts that are
powerful enough to yield dense, observable CSM -- and it is important
to know why. Is the relevant instability only significant at higher
initial mass and luminosity ( \S 3, 4.1, 4.2)?  Does the fraction of
SNe~IIn reflect rare interactions in a binary system at the right
separation (e.g. \S 4.3), or some other special circumstance?  The
process is rare, but not too rare.  Do stars of lower initial mass
(say 10-30 \sol) that make the majority of SNe also encounter similar
instabilities before core collapse, but they are less powerful and do
not cause hydrodynamic ejection in these stars?  Is there a continuum
in energy and mass ejection in pre-SN eruptions that extends below
what can readily be detected in bright SNe~IIn, so that some fraction
of ``normal'' SNe may also suffer pre-SN instability?  If the pre-SN
activity is too weak to cause severe mass ejection, does it
nevertheless have potential implications for the initial conditions
for core collapse (\S 4.4)?  Although these questions are inspired by
observations and get to the heart of the puzzle, observations alone do
not yet yield definitive answers.  Below we discuss these issues from
a theoretical perspective.

\section{Insights from 3D simulations of Turbulence in Late Stages}

In parallel to the observational progress outlined in \S2, theoretical
progress, in the form of direct numerical simulations of 2D and 3D
turbulence in late burning stages of stars, has provided valuable
insights into the physical nature of this evolution
\citep{ba94,ba98,aa00,ma06,ma07b,am11a,maxime2}.  These simulations
indicate that 1D models, as presently implemented, will not capture
essential aspects of turbulent convection (especially its
  time-dependent and fluctuating nature) or its influence on stellar
structure and evolution.  This, in turn, may lead to inaccuracies in
pre-SN stellar structure that are fundamentally important.

In what follows we will use some common terms in a precise way;
  for clarity we define them here.
\begin{itemize}

\item perturbation: a change in a variable which is small compared to
  its average value, so that linear perturbation theory is valid.
\item fluctuation: a change in a variable which is comparble to its
  average value, so that nonlinear terms are important, and linear
  analysis is invalid.
\item eruption: an event in which the rate of change of the average
  value of a variable (the ``background'') is comparable to that of
  fluctuations. Such events are often associated with mass loss
  because, by this definition, the energy of the eruption is
  comparable to the internal energy, and hence the binding energy of the star.
\end{itemize}
In the stellar context, eruptions often lead to mass loss.  They may
be especially violent and significant in stars near the Eddington
limit with loosely bound envelopes.

\subsection{An Example of O-burning in 3D}

Significant progress has been made in understanding such
  turbulent, dynamic behavior theoretically, by careful analysis of
  well-resolved 3D numerical simulations.  \cite{ma07b} examined the
  burning of $^{16}\rm O$ through 8 turnover times in a $23 \msol$
  star, a few days prior to core collapse (as estimated by 1D models). 
  There was no other significant burning on the grid, and the
  initial state was early in oxygen shell burning.  The initial
  evolution in this 3D simulation was a rapid readjustment
  in which a dynamically self-consistent set of pressure and velocity fluctuations
  grew from low level noise (turbulent kinetic energy increased from less than
  $10^{42}$ to more than $10^{47}$ ergs) to form a convective region. 
  The lack of consistent turbulent fluctuations is a
  common characteristic of 1D models: they are not self-consistent
  regarding the dynamics of convection. 
  Even well chosen guesses at initial velocities do not help because they must be
  properly phased relative to entropy fluctuations, 
  information we do not have from 1D models. The natural state
  of the convection, which develops in another turnover time, is
  dynamic, and structured in a complex way in {\em both} space and time,
  as demanded by the fluid flow equations in the turbulent regime.

  As the simulation in \cite{ma07b} advanced, the convection proceeded
  with pulses of turbulent kinetic energy. The burning was mild,
  almost constant, and the fluctuations in kinetic energy and
  luminosity instead came from the turbulent flow.  \cite{am11b} show
  that this dynamic behavior is similar to that generated in the
  \cite{lorenz} model of a convective roll, well known in meteorology
  and dynamic systems theory. During this early, mild stage of oxygen
  burning, the fluctuations did not (yet) induce significant changes
  in the energy generation rate, although fluctuations in turbulent kinetic
  energy and in enthalpy flux give variations of more that 50\%.

\subsection{Simulating A More Extreme Case}

Of particular relevance are the subsequent results of \cite{am11a} in
which it was shown that 2D simulations, of combined O and Si burning
shells in a 23 \sol \ star, evolved from a quasi-steady state into
eruptive behavior. The eruptions are violent and probably cause
extensive mass loss, but do not blast the core completely apart. 
  In this more extreme case the fluctuations in temperature induced by
  turbulent flow do affect the energy generation rate, and allow
  coupling between the flow, the oxygen burning, and the silicon
  burning. This is a new and complex problem, and the time-dependence
  of burning may be of fundamental importance to observed transient
  sources.  The prediction of this eruptive behavior is surprising,
being entirely different from that inferred from 1D stellar evolution
codes, for which O and Si burning stages are assumed to be
quasi-static.  However, indications of this violent behavior have been
seen in all hydrodynamic simulations of Si burning which could show it
\citep{ba98,meakin06,am11b}.  Another example of violent,
nondisruptive behavior is the pulsational electron-positron pair
instability SN models mentioned in \S\ref{epair}, but they are
extremely rare because they will be limited to the massive cores
arising from stars with extremely high initial mass.  Unsteady nuclear
burning may be spread across a wider range of initial mass.

\subsection{The importance of fluctuations}

The calculations reviewed above illustrate that there is an important
limitation in the standard 1D approach to the late stages of stellar
evolution: it ignores finite amplitude fluctuations which have
nonlinear interactions with the nuclear burning.  Here {\em
  fluctuations} differ from {\em perturbations} in that the nonlinear
terms are {\em not} small enough to discard, so the assumption of
  quasi-static burning is invalid.

Linear stability theory (e.g., \citealt{mbh04}) will not capture this
behavior because the driving and damping terms are nonlinear (they
involve  second- and third-order correlations between fluctuations; 
  \citealt{am11b} ).  The problem gets worse (the nonlinear effects
get bigger) as neutrino emission accelerates the evolution of the
star 
 in the final moments approaching core collapse \citep{fh64}.
  \cite{unno89} have stressed that the standard version of stability
  theory now used is missing a potentially important term: the dynamic
  effects of convection, which is on the same level as the terms
  giving rise to the excitation mechanisms for the standard
  instabilities (e.g., the $\epsilon$, $\kappa$, and $\delta$
  mechanisms, {\it op. cit.}, p. 241).

 \cite{am11b} have derived a turbulence model as a possible
  replacement for mixing length theory (MLT), with the constraint that
  it reproduce the dynamic, fluctuating behavior seen in the 3D
  simulations \citep{ma07b}. It includes two advances which were
  published in the western literature only after the original work of
  Erika B\"ohm-Vitense \citep{bv58}: the Lorenz model \citep{lorenz}
  and the Kolmagorov damping at small scales \citep{kolmg}; see
  \cite{am11b} for details.   We will re-derive the B\"ohm-Vitense
equation of mixing length theory, upon which most\footnote{The theory
  of \cite{cm91} makes equivalent approximations with regard to
  fluctuations, so that our discussion is applicable to this class of
  codes as well.} stellar evolutionary codes are based, and we will
show that it artificially suppresses fluctuations.
This  illustrates, explicitly and analytically, that conventional
stellar evolution ignores a major dynamical aspect of late stages of
stellar evolution --- the same stages during which the existence of
eruptions and explosive mass loss now have firm observational evidence
(see \S2),  as well as confirmation in the multi-dimensional
  fluid-dynamic simulations \citep{am11a}.

\subsection{MLT and the Lorenz model}\label{vel-scale-sect}

It has long been known that nothing like an MLT blob appears in
multi-dimensional well-resolved simulations (e.g., \cite{fls96},
\cite{sn98}, \cite{ba98}), yet MLT has proven so useful that it is
still the algorithm of choice for most stellar evolution codes (see
Langer 2012 for a review).

Perhaps there is some truth in the {\em mathematics} of the MLT
equations, if not the {\em physics} of the blob picture. If so, then
it should be possible to derive the same (or very similar) equations
from a conceptual picture more consistent with the behavior actually
seen in the simulations. We do this below.

The mathematical basis for MLT as used in stellar evolution is the
B\"ohm-Vitense cubic equation \citep{vitense53,bv58} which may be
written as
\begin{equation}
x^3 + {8 \over 9} U (2Ux + x^2-W),\label{bvcubic}
\end{equation}
where $x = \sqrt{(\nabla - \nabla_e)}$ and $ W = \nabla_r - \nabla_a$
(see \cite{kippen}, \S7.2 for this notation). This equation is
constructed from three equations (their Eq.s 7.6, 7.14 and 7.15),
which are
\begin{eqnarray}
v^2 = u_0^2 (\nabla - \nabla_e) { \ell_m^2 \over 8 H_P^2} \label{MLT1}\\
\nabla_e - \nabla_a = 2U \sqrt{\nabla - \nabla_e}\label{MLT2}\\
(\nabla - \nabla_e )^{3 \over 2} = {8 \over 9}U ( \nabla_r - \nabla )\label{MLT3}.
\end{eqnarray}
Here $u_0^2 = g \beta_T H_P$, and
\begin{equation}
  U = {3acT^3 \over C_P \rho^2 \kappa \ell_m^2} \sqrt{{8H_P \over g \beta_T}} , \label{BV1}
\end{equation}
where $H_P$ is the pressure scale height, $\ell_m$ the mixing length,
$\beta_T$ the coeficient of thermal expansion, $C_P$ the specific heat
at constant pressue, and the other symbols have their usual meaning.
The expressions $\nabla$, $\nabla_e$, $\nabla_a$, and $\nabla_r$
represent the dimensionless temperature gradients ($(\partial \ln
T/\partial \ln P)_i)$, for the background, the convective element, the
adiabat, and that temperature gradient required to carry all the
luminosity, respectively.

\begin{figure}
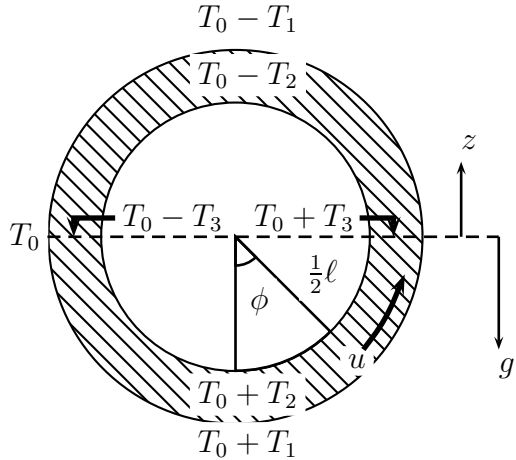

\figurenum{1}
\
\psset{unit=.5cm}
\pspicture*[](-12,-2)(4,12)
\pscircle[fillstyle=vlines](-5,5){5}
\pscircle[fillstyle=solid](-5,5){3.6}
\psline[linewidth=1pt]{<-}(2,2)(2,5)
\rput*[l]{N}(2,1.5){\large \bf $g$}
\psline[linewidth=1pt]{->}(1,5)(1,7)
\rput*[l]{N}(1,7.5){\large \bf $z$}
\rput*[l]{N}(-11,5){\large \bf $T_0$}
\rput*[l]{N}(-8,5.5){\large \bf $T_0-T_3$}
\psline[linewidth=2pt]{->}(-8.2,5.5)(-9.3,5.5)(-9.3,5.0)
\rput*[l]{N}(-4.5,5.5){\large \bf $T_0+T_3$}
\psline[linewidth=2pt]{->}(-1.7,5.5)(-0.8,5.5)(-0.8,5.0)
\pswedge[linewidth=1pt,fillcolor=lightgray](-5,5){3.6}{270}{315}
\rput*[l]{N}(-4.6,3.3){\large \bf $\phi$}
\pswedge[linewidth=1pt,fillcolor=lightgray](-5,5){0.8}{270}{315}
\rput*[l]{N}(-3.1,4.0){\large \bf ${1 \over 2}\ell$}
\rput*[l]{N}(-6,10.8){\large \bf $T_0-T_1$}
\rput*[l]{N}(-6,9.3){\large \bf $T_0-T_2$}
\rput*[l]{N}(-6,-0.5){\large \bf $T_0+T_1$}
\rput*[l]{N}(-6,0.8){\large \bf $T_0+T_2$}
\rput*[l]{N}(-2,1.7){\large \bf $u$}
\psline[linewidth=2pt,linearc=4.5]{->}(-1.6,2)(-1,2.8)(-0.5,4)
\psline[linewidth=1pt,linestyle=dashed]{-}(-10,5)(2,5)
\endpspicture
\caption{The Lorenz Model of Convection: Convection in a Loop (after
  \cite{am11b}). The speed of flow around the loop is $u$, and $T_i$,
  $i=1,2,3,4$ represents the amplitude of the potential temperature
  fluctuation relative to $T_0$, at different key points in and around
  the loop. 
  }
\label{lorenz}
\end{figure}
\placefigure{1}

\subsection{The Vortex Model}

MLT is not unique; we illustrate this with another theoretical model
that can produce  a cubic equation like that of B\"ohm-Vitense
  (Eq.~\ref{bvcubic}). To construct an alternative, we begin with the
\cite{lorenz} model of a convective roll (see Fig.~\ref{lorenz}),
which has been shown to exhibit key properties of the simulation
behavior as noted above \citep{amy09,am11b}.  This model includes
chaotic behavior and finite fluctuations in velocity and luminosity.
The dissipation is assumed to be that implied by the Kolmogorov
cascade, which affects the effective Peclet number\footnote{The Peclet
  number is the ratio of the advective transport rate to the diffusive
  transport rate. In a turbulent medium its definition requires some
  care because of the turbulent cascade.} for the flow \citep{am11b}.
Transformation of variables for non-dimensionality leads to
\begin{eqnarray}
\tau &=& t\ K\nonumber \\
X&=&u\ (2 / \ell K)\nonumber\\
Y&=&T_3\ (g/2 \ell \Gamma K T_1 )\nonumber\\
Z&=&T_4\ (g/2 \ell \Gamma K T_1)  ,
\end{eqnarray}
where $\sigma= \Gamma/K$ is the Prandtl number\footnote{The Prandtl number is the ratio of
the momentum diffusivity to the thermal diffusivity.}, $\Gamma$ is the inverse of 
the effective viscous damping time, $K=\nu_T(2/\ell)^2$ is the thermal relaxation time,
$\nu_T$ is the thermal diffusivity,
$t$ is the time variable, 
$\ell$ is the diameter of the convective roll, and $u$ is the convective speed around
the roll. The aspect ratio of the roll is $a$ so that $b=4/(1+a^2)$
deals with the excess in vertical over horizontal heat conduction. Lorenz took $b=8/3$
and $\sigma=10$. These values give behavior similar to that found in 3D simulations; see \cite{am11b} for further discussion.

Fig.~\ref{lorenz} illustrates the temperatures around the roll. 
The mean temperature at the midplane of the roll is $T_0$; the amplitude of the variation
in background temperature in the vertical direction is $T_1$. The fluctuating temperature
amplitude in the horizontal direction is $T_3$; in the vertical it is $T_2$. It is convenient to
define the temperature amplitude difference $T_4 = T_1 - T_2$. These are ``potential temperatures'':
variations relative to the adiabatic run of temperature.

We have
\begin{eqnarray}
dX/d\tau =& - X\vert X \vert/2 + \sigma Y    \label{lor1b}\\
dY/d\tau =& -XZ +rX -Y \label{lor2b}\\
dZ/d\tau =& XY - bZ, \label{lor3b}
\end{eqnarray}
where $r = (g \ell / 2 C_P T_1)\sigma$ is the Rayleigh number\footnote{The Rayleigh number indicates the onset of buoyant convection.}.  The
only functional change with respect to the original set of equations
\citep{lorenz} is the dissipation term in the $dX/d \tau$ equation,
with $\sigma X \rightarrow X \vert X \vert/2$. This is a consequence of the fact that, 
for stellar dimensions, viscosity is effective at
the end of the turbulent cascade, not at the integral scale of the roll.

Eq.s \ref{lor1b}, \ref{lor2b} and \ref{lor3b} define a ``vortex''
model for convection if we use the identifications made below for
stellar variables. Numerical solution of these equations in the
context of hydrodynamic stellar structure gives rise to chaotic
behavior similar to that found by \cite{lorenz}; see \cite{am11b}.

\subsection{The Steady Vortex Model}
 
To make contact with MLT, we take the steady state solutions of the
vortex model; we will call this the ``steady vortex'' model. We then
have three algebraic equations:
\begin{eqnarray} 
X^2 =& 2 \sigma Y, \label{sslor1}\\
rX=& XZ + Y,\label{sslor2}\\
XY=&bZ\label{sslor3},
\end{eqnarray} 
as in \cite{lorenz} for his steady state case.  In MLT $X\vert X
  \vert \rightarrow X^2$ by definition. Converting back to
dimensional variables (see Fig.~\ref{lorenz} and \citealt{am11b}), the
first equation becomes
\begin{equation}
u^2 = {\ell_d \over 2} g \beta_T {T_3 \over T_0},\label{L1}
\end{equation}
where the Kolmogorov damping length is $\ell_d = \alpha_d \ell$ and
the diameter of the roll is $\ell$.

Unlike MLT, $\alpha_d$ is {\em not} an adjustable parameter, but
  may be determined directly from simulations: for a shallow
convection zone $\alpha_d \approx 0.8$, while for a deep one (strong
stratification), $\alpha_d \approx 0.4$ \citep{maxime2}.  The $T$
variables are temperature amplitudes and $u$ is the velocity amplitude
as defined by \cite{am11b}. We use potential temperature (temperature
relative to its adiabatic value) to include some effects of
compressibility and stratification (we emphasize that this includes
some effects, not all; see \cite{maxime2}).

The second dimensionless equation becomes
\begin{equation}
u {T_2 \over T_0} = \nu_T {2 \over \ell} {T_3 \over T_0}, \label{L2}
\end{equation}
where 
\begin{equation}
\nu_T = {4aT^3 \over \rho C_P }{c \over 3 \rho \kappa}. 
\end{equation} 
To keep the algebra concise, we define
\begin{equation}
\bar{U} =  \Big ( {2 H_P \over  \ell_M} \Big )^2 { \nu_T \over H_P  u_0},
\end{equation}
which is similar to the mixing length quantity $U$ defined above,
\begin{equation}
U = {9\sqrt{2}\over 8} \bar{U}.
\end{equation}

The third equation becomes
\begin{equation}
u {T_3 \over T_0} = b \nu_T {2 \over \ell} {T_1 - T_2\over  T_0}. \label{L3}
\end{equation}

In order to once again make contact with MLT variables, we equate
finite differences of amplitudes to derivatives.  The temperature
  field in the vortex model is two-dimensional (horizontal and
  vertical), not one-dimensional (radial) as in MLT.  The horizontal
  derivative in density (temperature) is needed to drive a buoyant
  torque, so we identify
\begin{equation}
{T_3 \over T_0} \Big/ {\ell \over 2 } = (\nabla      - \nabla_e)/H_P , \label{TR1}
\end{equation}
and Eq.~\ref{L1} becomes
\begin{equation}
u^2 = u_0^2   (\nabla - \nabla_e) { \alpha_d \ell^2\over 4H_P^2},\label{321D1}
\end{equation}
which is to be compared to Eq.~\ref{MLT1}.  With the appropriate
choice of mixing length $\ell_m$ they are identical. This implies
 \begin{equation}
\ell_m = \sqrt{2 \alpha_d}\ell \approx \sqrt{1.6}\ell \approx 1.26 \ell. \label{ell1}  
\end{equation}
In the Lorenz model, $\ell$ has a precise meaning: the diameter of the
roll, whereas in MLT $\ell_m$ is an adjustable parameter of order
$H_P$. Here $u_0$ appears as a natural velocity scale, where $u_0^2 =
g \beta_T H_P$.

 We further identify the vertical temperature gradient relative to
  the adiabatic gradient by
\begin{equation}
{T_2 \over T_0} \Big/ {\ell \over 2 } =(\nabla_e - \nabla_a)/H_P , \label{TR2}\\
\end{equation}
and finally the imposed vertical gradient relative to adiabatic as
\begin{equation}
{T_1 \over T_0} \Big/ {\ell \over 2 } = (\nabla_r  - \nabla_a)/H_P.   \label{TR3}
\end{equation}
Using Eqs.~\ref{TR1} and \ref{TR2} with \ref{L2} gives
\begin{equation}
\nabla_e - \nabla_a = 2\bar{U} \sqrt{\nabla - \nabla_e}, \label{321D2}
\end{equation}
which is to be compared to Eq.~\ref{MLT2}.  The two factors have many
components in common; equating them implies
\begin{equation}
\ell_m = {3\over 2}(2\alpha_d)^{1\over 4} \ell \approx 1.7 \ell, \label{ell2}
\end{equation}
which relates the MLT mixing length to the diameter of the convective
roll.  This differs from Eq.~\ref{ell1}, but not drastically.  This
contradiction stems from the fact that in MLT, one adopts the length
of a blob mean-free-path as being the same as its size for radiative
cooling \citep{amy10}; this is related to the ``geometric factor'' problem.

 Using Eqs.~\ref{TR1}, \ref{TR2}, and \ref{TR3} with \ref{L3} gives
\begin{equation}
(\nabla - \nabla_e)^{3 \over 2} =b\ 2 \bar{U}  (\nabla_r - \nabla_e ), \label{321D3}
\end{equation}
which is to be compared to Eq.~\ref{MLT3}.  There are two differences
in these two equations. First, Eq.~\ref{MLT3} has a factor $\nabla_r -
\nabla$ while Eq.~\ref{321D3} has an apparently different factor
$\nabla_r-\nabla_e$.  This equation arises from the condition that the
total flux is the sum of convective and radiative fluxes (flux of
turbulent kinetic energy is ignored in both cases).  Both theoretical
models require this to be true, but define the radiative flux
differently in the convective zone.  Outside the convective region,
$\nabla_r = \nabla =\nabla_e$, so there is no difference. Inside the
convective region, the Lorenz model takes the whole region to be the
convective ``element'' (this resembles the simulations), so that the
radiative flux is that of the ``element''. MLT assumes that the
``element'' is distinct from the background, and that the radiative
flux is that of the background. This difference is large only in the
superadiabatic region, where neither theory resembles simulations nor
is self-consistent (see above).  In most stars the thickness of the
superadiabatic region is determined by the hydrogen recombination
region, and is much less than a pressure scale height. Such details
are ignored in both MLT and the vortex model, but they are prominent
in simulations \citep{sn98,nsa}.

  The second difference is in the coefficients multiplying these
  factors, equating them gives
\begin{equation}
{8\over 9}U =
b \  2\bar{U}
\end{equation}
Removing the common factors gives
\begin{equation}
\ell_M = {1 \over b}(2\alpha_d)^{1 \over 4} \ell \approx {3 \over 8} (1.6)^{1\over 4} 
\ell \approx 0.422 \ell. \label{ell3}
\end{equation}
While the choices (Eq.~\ref{ell1}, \ref{ell2}, and \ref{ell3}) for
mixing length needed for equality are surprisingly similar, they are
not identical due to the different conceptual foundations used. Note
that Equations \ref{ell1} and \ref{ell2} are necessarily inconsistent
if we make the reasonable choice of using the $\alpha_d$ determined by
simulations. At this level of precision MLT is inconsistent with
hydrodynamics.

The Lorenz model is mathematically precise for the single mode of
  flow which was chosen; simulations show that several low-order modes
  ($\sim 5$) dominate, so that a single mode picture is
  oversimplified; e.g., see \cite{am11b}. The MLT derivation has a
  plethora of factors of 2 and of geometry which are not unique, and
  MLT has adjustable parameters. It is unlikely that this difference
  in factors (of order unity) is significant.

Both MLT and Lorenz convection assume symmetry between up-flows and
down-flows; this is inconsistent with simulations, which due to
up-down asymmetry, have non-zero acoustic ($\bf F_P$) and kinetic
energy ($\bf F_K$) fluxes \citep{ma07b,amy09,maxime2}. The
  asymmetry increases for increasing stratification.

\subsection{B\"ohm-Vitense Equation Recovered}
 
As we have seen, the steady vortex model  produces three basic
  algebraic equations almost identical to those making up the
B\"ohm-Vitense version of MLT.  Combining Eq.s \ref{321D1},
\ref{321D2}, and \ref{321D3} gives an equation strikingly akin to
Eq. \ref{bvcubic} (the B\"ohm-Vitense cubic equation),
\begin{equation}
x^3 + 2\bar{U} (2\bar{U}x - W) = 0. \label{bvcubic2}
\end{equation}
These differ in (1) $U$ and $\bar{U}$ as discussed above, and (2) the
lack of an $x^2$ term due to the different definition of radiative
flux in a convective zone.  The $x^2 = \nabla - \nabla_e$ term
  may be neglected in both the large $x$ and small $x$ limits, and has
  only a modest effect in the transition region.  Eq.s~\ref{bvcubic}
  and \ref{bvcubic2} have the same asymptotic behavior and similar
  constant factors. Physically this transition between large and small
  $x$ corresponds to that between the adiabatic gradient and the
  radiative gradient. The central regions of massive stars are
  never in this transition region; it occurs in the Sun (and other stars) near the
  surface, associated with the hydrogen ionization zone.  {\em On
  purely mathematical grounds, the steady vortex model must give
  comparable agreement with observations as MLT.} Inside a stellar
evolutionary code the two are almost identical, and any small
differences may be removed by slight adjustments of assumptions
involving flow patterns, or mixing length and geometric parameters,
for example.
 
%
%

The mathematical equivalence of MLT and the steady vortex model via
the B\"ohm-Vitense equation should serve as a warning. The physical
pictures used in the two models are different, and therefore the
interpretation of MLT in terms of physical processes is not
unique. The steady vortex picture may prove more useful for physical
interpretation, to the extent that it has connections to actual
solutions of the Navier-Stokes equations in the form of 3D numerical
simulations.
 
{\em Both MLT and the steady vortex picture lack fluctuations.  For
  the steady vortex picture this is because use of the steady state
  solutions removes the terms that cause  chaotic behavior (the
    Lorenz strange attractor). These terms do not exist in MLT.}

Thus, Eq.s~\ref{sslor1}, \ref{sslor2}, and \ref{sslor3} (the {\it
  steady} vortex model) have no  chaotic behavior while
Eq.s~\ref{lor1b}, \ref{lor2b}, and \ref{lor3b} (the {\it dynamic}
vortex model) do.  This explicitly demonstrates a fundamental
difference between MLT and the more general dynamic vortex model. It
is equivalent to using a damping (to get the steady state solution),
and ignoring the dynamic aspects of turbulence.  This neglect has
potentially important consequences for models of the late stages of
stellar evolution, which we will discuss below.
 
The dynamic vortex model, which is more directly based on the 3D fluid
dynamics equations than MLT, allows a more direct way to generalize
stellar evolution codes by including additional physics, and a better
conceptual base to deal with the physics involved.  It can be shown
that use of an acceleration equation in a vortex picture provides a
way to include dynamics, boundary layers, and the turbulent cascade
into a simple convection model (Arnett, et al., in preparation). In
addition, it provides a natural link to the Richardson-Kolmogorov
cascade and gives a balance between driving and damping of turbulent
flow.

 Stellar evolutionary models are based on an implicit assumption
  of stability.  The most common tool for exploring instability is
  linear perturbation analysis \citep{cox80,unno89}, which deals with
  convection poorly (although perhaps adequately for the relatively
  stable main-sequence phase). The driving term for turbulence is
  quadratic and the damping term is cubic; neither is linear, so
  linear analysis is blind to them. However, numerical hydrodynamics
  is not so blind, and such simulations clearly deviate from stellar
  evolution models in the latest stages of a massive star's life, for
  which turbulent convection plays a dominant role. In this paper we
  have precisely identified the culprit --- the chaotic behavior
  implicit in a convective roll.  Steady-state solutions become
  irrelevant when they are unstable; the \cite{lorenz} model of a
  convective roll, with its stange attractor (chaotic behavior),
  illustrates this. It appears that the 3D simulations have similar
  chaotic behavior which give rise to convective pulses \citep{ma07b}
  and perhaps eruptions \citep{am11a}.
 
  Why not just simulate the whole star? The combination of the need
  for high resolution to capture the turbulence (high numerical
  Reynolds number is required) and sufficient volume to include both
  the core and active burning shells, results in a computational
  problem that strains present-day computer facilities. This is
  presently being attempted (Meakin \& Arnett, in preparation), but it
  is appropriate now to examine the implications of this new
  perspective.

\section{Implications}
 
In order for models to provide a coherent picture of the observations,
we need some process to be able to inject $10^{48}$ to $10^{50}$
ergs\footnote{This range of energy comes from the most extreme events
  ($\eta$ Car, SN~2006gy's precursor eruption, etc.) to the least
  extreme LBV-like giant eruptions (such as V12 in NGC~2403,
  SN~2000ch, precursors of SN~2006jc and SN~2009ip), plus a number of
  objects in between. These exhibit a range of $\sim$100 in ejected
  mass at similar outflow speeds (see Smith et al.\ 2011b).  Note that
  individual eruptions with energy well below 10$^{48}$ ergs may
  exist, but could be difficult to detect in external galaxies.  On
  the other hand, energy injection of 10$^{47}$ erg or less may be
  absorbed by the stellar envelope, and might not lead to the
  hydrodynamic ejection that gives rise to a brief optical transient
  anyway (e.g., Dessart et al.\ 2010).} in the final couple of years
before core collapse (i.e. during Ne, O, and Si burning), and we may
need something similar but with lower energy to be more prolonged
(several hundred years) during C burning and possibly even He burning.
While Ne burning is weak and Si burning complex, both C and O burning
release about $ 5 \times 10^{17}$ ergs g$^{-1}$ of fuel burned (or
equivalently, $10^{51}$ ergs per $M_{\odot}$ of fuel burned), so that
a modest amount of nuclear burning could directly provide the energy
needed. This is relevant to the issue of shell instabilities
\citep{am11a}.

Must the energy be supplied directly, or could it be the result of
some less than completely efficient process?  For core collapse to
occur, at least $1.5 \times 10^{51}$ ergs (1.5 bethes) must be
released by nuclear burning to convert helium burning products into
iron-peak nuclei\footnote{A mass of 1.5 $M_\odot$ converted from
  C$^{12}$ and O$^{16}$ releases about $5 \times 10^{17}$ ergs
  g$^{-1}$ in burning to Si$^{28}$ \citep{fh64}, or $5 \times 10^{17}
  \times 3 \times 10^{33} = 1.5 \times 10^{51}$ ergs.  Further burning
  could release slightly more energy, but is countered by neutrino
  loss from nuclear weak interactions \citep{wda96}.}.  This is much
larger than the gravitational binding energy of the star, so that {\em
  to avoid explosion and mass loss} this energy must be radiated away.

{Neutrinos can do the job, but vigorous convection is demanded. It
  has long been known \citep{fh64,wda96} that, because of the more
  sensitive temperature dependence of thermonuclear heating in
  comparison to neutrino cooling, that a convection zone in thermal
  balance will have heating at the bottom and cooling at the top. This
  generates an entropy gradient that drives convective transport of
  heat upward.
  To prepare the core for collapse requires turbulent convection to
transport 1.5 bethes of energy, a factor of 15 more than the largest
value needed to explain the observed pre-collapse ejection of
mass. Any inefficiency in the transport process can therefore
contribute to the injection of energy into a star's outer envelope.
Depending on how much energy is diverted, the results can range from
modest to catastrophic.  It would take less than 7\% of 1.5 bethes to
supply the largest eruption energy ($10^{50}$ ergs) required above.

Below we highlight six ways in which the dynamical instability
associated with turbulent convection in late burning stages might
profoundly influence the star's pre-SN evolution, the stellar
structure at the moment of collapse, the observational interpretation
of supernova remnants, and isotopic abundances in meteoritic pre-solar
grains.

\subsection{Enhanced mass loss}

A characteristic of turbulence is the presence of chaotic fluctuations
in luminosity and velocity. 
Convective velocities rise from $\sim$0.5 km s$^{-1}$ in He burning to
100 km s$^{-1}$ in O burning, and 300 km s$^{-1}$ in Si burning
\citep{ma07b,am11a}.  This is subsonic in the core (mach number $\sim
0.01$ to $0.1$) but not in the envelope.  Simulations show vigorous
wave production at convective boundaries \citep{ma07b}.  Waves
generated by vigorous turbulence travel outward and dissipate (e.g.,
steepen into shocks); consequently, the envelope becomes an absorber
for this energy. If the envelope is already near Eddington luminosity,
it can only rid itself of this extra energy by dynamical expansion or
mass loss.  For red giant structure, this would happen on a pulsation
(sound travel) time scale, and appear as an eruption. For more
condensed structure (a star stripped of its H envelope, for example),
it might be seen as a vigorous wind\footnote{\cite{kk13} have very
  recently argued that the features of some strange SNe could be
  explained by energy input into the oxygen shell during Si burning.}.

Quataert \& Shiode (2012) have discussed this mechanism of enhanced
pre-SN mass loss driven by waves propagating through the star's
envelope as a result of furious convection during O and Ne burning.
This type of mechanism has the potential to explain strong winds that
may produce dense CSM in the few years leading up to core collapse,
although it is unclear if it can account for the most violent
explosive events that occur.

\subsection{Unsteady or explosive burning}

The episodic nature of the observed pre-SN mass loss suggests that, in
addition to strong winds that are quasi-steady state by definition, we
should consider hydrodynamic instabilities in shell burning with
energy injection by explosive/unsteady nuclear burning.  Both would be
in operation once neutrino cooling dominates over photon cooling, that
is, in the months and centuries prior to core collapse, from C burning
to Si burning.  There should be a wide variety of behavior because
both the instabilities and the wave transport will be different for
different core-envelope structures and rotational rates, both of which
also depend upon the history of any binary interaction.

To make a bomb requires an energy source and an ignition mechanism to
release that energy. The 1.5 bethes which are available from burning
the ashes of He burning to iron-peak nuclei is a more than ample
source; only a small fraction is required.  For an ignition mechanism
we already have the instability found by \cite{am11a}. That simulation
was terminated because the matter was flowing off the limited grid,
but already had violently excited the lowest order mode available to
fluctuations, with every indication of continued growth. Mild shocks
were already forming. Despite the severe challenge to computer
resources, it is important to repeat these simulations in 3D, on a
$4\pi$ steradian grid, extended to later times (Meakin, \etal, in
preparation). It is also important that the hydrodynamic algorithms
used be non-damping; although anelastic, low Mach-number and implicit
methods may allow larger time steps \citep{maxime1,alm10}, they must
be validated to insure that they give negligable artificial
damping\footnote{This is a difficult and subtle problem; see
  \cite{bvz12,vasil13}.}  for this problem, and that they seamlessly
transition from turbulent to explosive flow.

There may be other instabilities to be found; the published library of
%
%
well-resolved simulations of pre-collapse evolution is still
small. While oxygen shell burning seems quasi-stable over 8
turnovers  \citep{ma07b}, it is vigorous; the possibility of
eruption after a longer time remains open. Both C and Ne burning are
more difficult to simulate directly because of their slower burning
(which implies that more time steps are needed to calculate them over
significant evolutionary times). We do not yet know if they harbor
nonlinear instabilites which would become evident over the time scales
needed to consume C and Ne.

\subsection{Triggering violent binary interactions}\label{violent_binary}

There is another avenue for extremely sudden and violent events to
occur.  Recent observations of main-sequence O-type stars have shown
definitively that the majority of massive stars (roughly 2/3) are born
in binary systems with a separation small enough that the two stars
will interact before they die (Sana et al.\ 2012; Chini et al.\ 2012;
Kiminki et al.\ 2012).  The fate of this interaction depends on the
initial separation: the closest systems will exchange mass or merge on
the main sequence, while binary systems with wider separations will
interact only when the more massive star evolves off the main sequence
and expands to fill its Roche lobe as a supergiant.

If the mechanisms mentioned above (unsteady or explosive burning,
waves generated by vigorous convection) are able to inject an amount
of energy into the star's envelope that is {\it not quite sufficient}
to completely unbind the envelope, the result may be a swelling of the
star's hydrostatic radius (i.e. a large pulsation).  In that case,
dramatic events may ensue if the star is in a binary system with an
orbital period ranging from 10s of days to a few yr (depending on
eccentricity).  If the SN progenitor increases its radius
significantly, a companion star that previously had been too distant
to interact may suddenly find itself to be the victim of mass
transfer, a merger, or a violent collision if it is in an eccentric
orbit.

Consider the case of $\eta$ Car, where something similar is known to
have occurred: $\eta$ Car is a binary with a 5.5 yr orbital period and
an eccentricity $e$=0.9-0.95.  Smith \& Frew (2011) showed that brief
$\sim$100 day brightening events occurred at times of periastron, and
Smith (2011) argued that a stellar collision must have occurred
because the periastron separation was much smaller than the required
photospheric radius at that time.  Mauerhan et al.\ (2013) pointed out
that the multiple brief peaks in the few years leading up to the 2012
explosion of SN~2009ip were reminiscent of $\eta$ Car's events, and
suggested that violent periastron encounters may play a role.  The SN
impostor SN~2000ch may be yet another example (Pastorello et al.\
2010; Smith et al.\ 2011b).  A related idea was suggested by Chevalier
(2012), who also included the intriguing possibility of mergers with a
neutron star companion leading to SNe IIn, although he found that such
events are probably too rare to account for the observed frequency of
SNe~IIn.  

The possibility of energy injection that leads to envelope inflation
might act to enhance the frequency of any such merger/collision
events, because the progenitor {\it suddenly} finds itself to have a
larger radius and is more able to interact with more widely separated
companion stars (here ``suddenly'' means a time comparable to the
orbital period).  More importantly, if the increase in stellar radius
is a result of energy injected during C, Ne, O, or Si burning phases,
it provides a physical reason to expect such merger or collision
events to be {\it synchronized} to within only a few
years\footnote{The actual duration of these phases of C, Ne, O, and Si
  burning might be modified by their dynamic behavior.} before core
collapse.

\subsection{Preparing the core structure for collapse}

Numerical simulations have yet to reliably produce successful SN
explosions in a self-consistent way
\citep{wda96,kit06,bur07,lieb01a,lieb01b,ramp02,buras03,thom03,lieb05,sumi05,fy07,lentz12}.

While debate continues over details involved in these calculations
\citep{janka}, a unifying characteristic is that all results are
highly sensitive to the core structure of the model progenitor.  For
example, this progenitor structure determines the rate at which mass
rains down onto the proto-neutron star after core collapse (see
\citealt{oo13}, and their discussion of ``compactness'';
\citealt{ugliano12}).  Real SNe are observed to explode nonetheless.
Thus, it seems prudent to ask: What if the fault lies not so much in
the details of the method of computing neutrino transport, the
geometry (2D vs. 3D), or other aspects of the explosion, {\it but in
  the structure of the progenitor itself?}

%

The progenitor models most commonly used by the groups doing core
collapse simulations are those of Woosley and collaborators (e.g.,
\cite{whw02,hlw00,ww95}), all of which use MLT.  There is not yet good
detailed agreement in the final stellar structure among the various
evolution groups (see \cite{gem12}, \cite{chilim13}, Bill Paxton,
private communication), even though they all use some version of MLT
(this may be related to differing mixing algorithms).

What effect might a proper treatment of convection and its associated
hydrodynamic instabilities have on the stellar structure of a SN
progenitor?  Observations dictate that at least in some stars, pre-SN
instabilities inject energy into the star's envelope that leads to
significantly enhanced mass loss or even explosive ejection.  This is
not predicted by evolution models (but is indicated by the
hydrodynamic simulations), yet {\it something} must be providing
substantial extra energy on short timescales before collapse.  One may
surmize that successful mass ejection prior to core collapse is most
likely to occur in more massive stars, since they are closer to the
Eddington luminosity and have more loosely bound envelopes.  What
would happen, then, in lower-mass progenitor stars if the same process
occurs, but where the injected energy is insufficient to cause
observable eruptive/explosive mass loss?  As in
\S\ref{violent_binary}, one might infer that the outer layers may
swell and that binary interaction might be more likely, or
alternatively, that the mass loss might be less vigorous and the
consequences less easily observed.

We will not know what the detailed changes will be until realistic 3D
simulations illustrate the strongly nonlinear behavior. Some general
trends are clear, however. For example, both the enhanced mass loss
and unsteady burning move matter out of the potential well,
changing\footnote{Larger central condensation means reduced density in
  the part of the mantle which will fall onto the core, reducing the
  subsequent rate of mass infall and allowing the shock to propagate
  outward more easily. This is why smaller core masses explode more
  easily: they have larger central condensation.}  the central
condensation of the core and making the progenitors easier to explode
by the neutrino-transport mechanism \citep{oo13,ugliano12}.  The
``compactness'' parameter of \cite{oo13} is essentially the
gravitational potential at the edge of the core; in hydrostatic
equilibrium it acts as a pivot point for the density profile, so that
higher central density, and lower compactness, is accompanied by lower
mantle density (see Fig.~10.4 and 10.7 in \citealt{wda96}).  The mapping
of initial (ZAMS) mass to mass of the core at collapse may also be
altered. This is a complex problem because of unaddressed issues in 1D
evolution relating to convective boundaries \citep{ma07b,maxime2} as
well as large uncertainties in the treatment of mass loss
\citep{ns14}.

Turbulent fluctuations in the collapsing cores will also break
  spherical symmetry. They will be a result of both instantaneous
  driving by current burning and historical flows during the approach
  to collapse.  Estimates for the amplitude and the form of these
  fluctuations are available.  In the most advanced published
  simulation \citep{am11a} there seems to be a complex interaction
  between O and Si burning shells, both wildly turbulent, driving
  pulsational modes of the core and mantle.  The fluctuations are a
  factor of 10 or more larger than perturbations used to test core
  collapse simulations, and are of global rather than local scale (see
  \citealt{fy07,wjm10,wjm13} for simulations with lower amplitude
  fluctuations). Such fluctuations may lead to enhanced energy flow,
  and to increasing neutrino luminosity, so they may make explosion
  more likely.  Since a preprint of our paper was posted, simulations
  have confirmed our basic suggestion; \cite{couch&ott} have explored
  just one aspect of this problem, with encouraging results. Using the
  \cite{am11a} data, they imposed a modest nonradial velocity
  (momentum conserving), corresponding to an initial convective flow
  in the mantle of their collapsing core. At 100 milliseconds after
  core collapse, this material passed through the stalled shock, and
  by 200 milliseconds the core is exploding.  This small modification
  of the velocity field was sufficient to change a dud into a
  successful explosion by allowing the revival of the shock, showing
  the critical importance of realistic initial models for core
  collapse.

%

\subsection{Mixing in Presupernovae:  Meteoritic Grains}

The unsteady burning and its resulting eruptions may significantly
change the mixing history of the progenitor. Convective boundaries are
not sacrosanct; there is mixing across the boundaries even in
quasistatic phases \citep{ma07b,mocak08,mocak09}.  Eruptions add to
this, and turbulent mixing is faster and different from the
``diffusive convection'' algorithm currently in use.

Isotopes (nuclei with the same proton number Z but different neutron
number N) are good records of thermonuclear history because they are
resistent to changes caused by chemistry (different Z). The
development of precise microprobes and the discovery of grains in
meteorites whose crystalization predated the solar system has given a
new window into nucleosynthesis.  The changed picture of presupernova
evolution sketched above should have profound implications for the
interpretation of isotopic anomalies in pre-solar grains in meteorites
(for example, see \citealt{pign13,trav99,lug01}).  Not only do we have
possible changes in the bulk character of the initial models and the
explosion mechanism to consider, but also new processes. First,
turbulent fluctuations involve not only thermodynamic fields (e.g.,
temperature or density) and velocity, but also composition. Each
turbulent burning zone will introduce a ``cosmic scatter" in the
detailed abundances; see Fig.~4 of \cite{ma07b}.  Second, advective
mixing often proceeds by ``plumes'', which allow matter from one level
(and characteristic temperature and composition) to punch through
other layers to finally mix in yet another level (with different
composition), without homogenizing intermediate levels. This
``layer-jumping" may be more likely as the advection becomes more and
more violent. There seem to be unresolved issues in isotopic anomaly
interpretation that this might help.  Third, these two new processes
would occur before the explosion shock from core collapse, which
ejects the mantle from the core. Thermonuclear processing during this
explosion would therefore be altered.

\subsection{Mixing before, during, and after the explosion: Young SNRs}

Most of the interpretation of the observed asymmetries and mixing in
young SN remnants has proceeded with the assumption that the explosion
itself drove the asymmetries, rather than them being already developed
prior to explosion.  For core collapse this presents a problem. To
form a collapsed core implies a radial compression factor of 100 or
so. Asymmetries grow on compression, but decrease on expansion (this
is the fundamental problem of inertial confinement fusion,
\citealt{lindl}).  Core asymmetries will tend to be made smaller (more
spherical) on expansion by a large factor as compared to pre-existing
asymmetries in the mantle.  In addition, the mantle asymmetries are
likely to be increased by local explosive burning.

Young core-collapse SN remnants may contain evidence for asymmetries
directly related to turbulent fluctuations in pre-collapse phases.
Tantalizing new maps of abundances in Cas~A are now available
\citep{fesen13,isen12} which beg for detailed comparison with
multidimensional simulations \citep{carola13}.  Pre-existing
  density fluctuations in the mantle seem indicated
  \citep{isen12}. Such fluctuations will be caused by the type of
  turbulent convection in the progenitor that we described above.
Advective turnover occurs on a timescale slightly longer than the
explosive time scale, so that mixing will be partial, and will tend to
preserve (in a distorted way) any layering in abundance.  The more
thoroughly burned matter will have higher entropy, and punch through
overlying layers in some regions. Any $^{56}$Ni produced will decay
and heat the plasma, giving a characteristic modification (Ni bubbles)
of the explosive outflow \citep{fesen13}.

\section{Conclusions}
 
Both observations of pre-SN mass loss and direct numerical simulations
of convection are discordant with predictions of conventional stellar
evolution codes for the late stages of massive star evolution. In this
paper, we have discussed the observed nature of the most dramatic
discrepancies (pre-SN eruptions), and we have proposed that the
problem in 1D stellar evolution codes lies in the treatment of
turbulent convection.  In particular, 1D codes ignore finite amplitude
fluctuations in velocity and temperature, and their nonlinear
interaction with nuclear burning.  For most of a star's life this is
probably a reasonable approximation (except perhaps near convective
boundaries) but in the latest phases of
evolution such fluctuations can become catastrophic in massive stars.
The fluctuations are not allowed to occur in conventional 1D codes
that impose steady state behavior, so their associated pre-SN eruptions are not
predicted.

In order to derive MLT from a more general hydrodynamic formulation,
the following assumptions are needed: (1) the damping is consistent
with the Kolmogorov cascade, (2) the large scale dynamics could be
approximated by a Lorenz convective roll, and (3) a steady-state
solution was appropriate.  While the first two are consistent with the
3D simulations and with experimental work on turbulent flows, the last
is certainly not. Use of MLT therefore underestimates dynamic behavior
by artificially damping fluctuations in velocity and temperature.

This may be adequate (or not) through H and He core burning, but the
traditional treatment of convection becomes increasingly unrealistic
in the late stages of stellar evolution, which are accelerated by
neutrino losses. This stage of cooling coincides with an evolutionary
phase for which we have strong observational evidence of vigorous
eruptive mass loss. In contrast, convection for photon-cooled stages
(H and He burning) is less strongly modified, although treatments of
boundary layers, composition gradients, and discontinuities remain
questionable.

Carbon burning is an interesting transitional stage; it is the first
stage to be cooled primarily by neutrino emission, and the vigor of
carbon burning depends upon the abundance of $\rm^{12}C$ resulting
from the previous helium burning \citep{wda72c}. The $\rm^{12}C$
abundance also depends on the algorithm used for convective mixing
\citep{wda72b}. The precise reaction rates remain elusive; see, e.g.,
\cite{gai13,bsc13}. For small values of $\rm^{12}C$ (massive cores),
carbon will burn nonconvectively (and feebly), followed by weak neon
burning (carbon burning produces $\rm^{20}Ne$, the fuel for neon
burning). For large values of $\rm^{12}C$, carbon burning will be
vigorous, lasting $\sim 3 \times 10^{11}$ seconds ($10^4$ years), to
be compared with oxygen burning which lasts for $\sim 9 \times 10^8$
seconds (30 years) \citep{wda74b}. Carbon burning occurs at a lower
temperature, hence has a lower rate of neutrino cooling to balance
than oxygen burning, and can thus last longer. These values are quoted
for helium cores, which are more relevant to SNIbc events.  While
oxygen burning may be directly simulated in 3D, {\em turbulent} carbon
burning, over time scales long enough for carbon exhaustion, remains beyond present computational
capability. The behavior of massive stellar cores during carbon
burning rests on 1D models and speculation at present, but
observations suggest there may be enhanced episodic mass loss during
this stage (see section \ref{earlier_mdot}).
 
We conclude that the use of MLT and ``diffusive convection'' are
likely to be a dominant cause of the substantial discrepancies between
1D stellar models and actual observed pre-supernova stars. Realistic
inclusion of turbulence in stellar codes is a frontier topic, as is
the inclusion of rotation, binary interaction, and eruptive mass
ejection, with which it interacts.  Strong empirical evidence for
eruptions and violent binary interaction occuring in the few years
before core collapse suggests a link to instability from turbulent
convection in the latest burning phases.  Both interpretation of
isotopic anomalies in pre-solar grains in meteorites, and initial
models for gravitational collapse should be strongly influenced by the
likely alteration of progenitor structure.
   
Finally, we note that if the surface layers of a SN progenitor star
are indeed modified prior to the impending explosion in a predictable
way, it now becomes conceivable to design an early warning system for
core collapse.  The key would be to recognize the observational
signature resulting from energy deposition caused by the type of
mechanisms discussed herein.  They could manifest as a sudden swelling
of the star, sudden onset of violent binary interaction, or eruptive
mass loss; less extreme cases may in principle be recognized by their
asteroseismological signatures.  Some of these eruptive effects have
been observed in stars that have not yet experienced core collapse
(such as nearby LBV stars that experienced eruptions in past centuries
or millenia), so the observational patterns of pre-SN outbursts need
to be studied in more detail.  Indentification of pre-SN stars would
be useful for detection of their neutrinos, gravitational waves, and
early stages of nearby SNe and core collapse.  Present and future
transient surveys should be examined with this in mind.

\begin{acknowledgements}

  This work was supported in part by NSF Award 1107445 at the
  University of Arizona. One of us (DA) wishes to thank the Aspen
  Center for Physics (ACP) and the Kavli Institute for Theoretical
  Physics (KITP) for their hospitality and support. We wish to thank
  Casey Meakin, Maxime Viallet, Christian Ott, Jeremiah Murphy, and
  Bill Paxton for helpful discussions.


\end{acknowledgements}

\eject

\clearpage

\end{document}